\documentclass[twoside,leqno,twocolumn]{article}

\usepackage[letterpaper]{geometry}

\usepackage{ltexpprt}
\usepackage{hyperref}
\usepackage{hyperref}
\usepackage{graphicx}
\usepackage{multirow}
\usepackage{multicol}
\usepackage{adjustbox}
\usepackage{amsmath, bm}
\usepackage{amsfonts}
\usepackage{stackengine}
\usepackage{titlesec}
\usepackage{mathtools}
\usepackage{bibentry}
\usepackage{float}
\usepackage{amssymb}
\usepackage{cleveref}
\usepackage{graphicx,subcaption}
\usepackage{booktabs} 
\usepackage{ragged2e,xspace}
\usepackage[font=footnotesize, singlelinecheck=off]{caption} 
\usepackage{tabularx}
\usepackage[table]{xcolor}
\usepackage[T1]{fontenc} 
\usepackage[utf8]{inputenc} 
\usepackage[english]{babel} 
\usepackage{mathtools, nccmath}
\usepackage{lipsum}
\usepackage{hyphenat}
\usepackage{lmodern} 
\usepackage[ruled,commentsnumbered]{algorithm2e}
\newenvironment{packed_enum}{
\begin{enumerate}
  \setlength{\itemsep}{1pt}
  \setlength{\parskip}{0pt}
  \setlength{\parsep}{0pt}
}{\end{enumerate}}

\definecolor{aliceblue}{rgb}{0.94, 0.97, 1.0}

\definecolor{apricot}{rgb}{0.98, 0.81, 0.69}

\definecolor{babypink}{rgb}{0.96, 0.76, 0.76}

\definecolor{beaublue}{rgb}{0.74, 0.83, 0.9}

\definecolor{bisque}{rgb}{1.0, 0.89, 0.77}

\definecolor{antiquewhite}{rgb}{0.98, 0.92, 0.84}

\definecolor{almond}{rgb}{0.94, 0.87, 0.8}

\definecolor{champagne}{rgb}{0.97, 0.91, 0.81}

\definecolor{vanilla}{rgb}{0.95, 0.9, 0.67}

\definecolor{anti-flashwhite}{rgb}{0.95, 0.95, 0.96}

\definecolor{bananamania}{rgb}{0.98, 0.91, 0.71}

\definecolor{camel}{rgb}{0.76, 0.6, 0.42}

\definecolor{burlywood}{rgb}{0.87, 0.72, 0.53}

\newcommand{\trace}{\mathop{\mathrm{tr}}}
\newcommand{\apr}{AlignGraph\xspace}
\newcommand{\aprone}{G-align-Single\xspace} 
\newcommand{\aprtwo}{G-align-Double\xspace} 
\newcommand{\aprthree}{Fermat-Double\xspace}

\renewcommand{\eqref}[1]{(\ref{#1})}

\newcommand{\degdist}{D.D}
\newcommand{\clcoef}{C.C}
\newcommand{\aso}{ASRT}
\newcommand{\tri}{TRI}
\newcommand{\wc}{WG.C}
\newcommand{\clc}{CL.C}

\newcommand{\fullversion}[2]{#2}

\makeatletter
\newcommand{\removelatexerror}{\let\@latex@error\@gobble}
\makeatother
\SetKwInput{KwInput}{Input}                
\SetKwInput{KwOutput}{Output}
\begin{document}

\newcommand\relatedversion{}
\renewcommand\relatedversion{\thanks{The full version of the paper can be accessed at \protect\url{https://arxiv.org/abs/1902.09310}}} 

\title{\Large AlignGraph: A Group of \protect\\ Generative Models for Graphs}
\author{Kimia Shayestehfard
\and Dana Brooks
\and Stratis Ioannidis\thanks{\scriptsize{\{kshayestehfard, brooks, ioannidis\}@ece.neu.edu, Electrical and Computer Engineering Department, Northeastern University, Boston, MA, USA.}}}

\date{}

\maketitle







\begin{abstract} It is challenging for generative models to learn a distribution over graphs  because of the lack of permutation invariance:  nodes may be ordered arbitrarily across graphs, and standard graph alignment is combinatorial and notoriously expensive. We propose AlignGraph, a group of generative models that combine fast and efficient  graph alignment methods with a family of deep generative models that are invariant to node permutations. Our experiments demonstrate that our framework successfully learns graph distributions, outperforming competitors by $25\% -560\%$ in relevant performance scores. \end{abstract}

\section{Introduction}
Graph generative models have  applications across domains like chemistry, neuroscience and engineering. Generative models  learn a distribution over graphs, and are used to subsequently  sample from this distribution. They can, e.g., be used to predict interfaces between proteins during drug design and discovery~\cite{do2019graph,li2018multi}, or to perform hypothesis testing and simulation for social networks, when collecting real graphs is difficult~\cite{leskovec2010kronecker,kim2011}.




Traditional generative models for graphs such as the Barab\'{a}si-Albert~\cite{barabasi1999emergence}, Erd\"{o}s-R\'{e}nyi~\cite{erd1959and}, and stochastic block models~\cite{snijders1997estimation} generate graphs with provable formal properties but which often lack realism. For example, the Erd\"{o}s-R\'{e}nyi model produces graphs with a light-tailed degree distribution~\cite{barabasi1999emergence,watts1998collective}, while the Barab\'{a}si-Albert model fails to generate graphs with a high clustering coefficient~\cite{fronczak2002higher}. 
Deep generative models such as variational autoencoders~\cite{kipf2016variational} and graph recurrent neural networks~\cite{you2018graphrnn,liao2019efficient} have shown great potential in learning distributions  from graph datasets, 
at greater fidelity than traditional models. 
%
However, learning a distribution of graphs over a dataset poses a significant challenge because of the \emph{lack of permutation invariance}, since  graph nodes may be subject to arbitrary {permutations} across graphs: the correspondence between nodes in different graph samples may be a priori unknown. 

This is a problem because state-of-the-art generative models, like the ones listed above, rely on latent node embeddings. Such embeddings vary drastically even under nearly isomorphic graphs~\cite{gritsenko2021graph}. In turn, this can hamper the fidelity of the graph generation process significantly. Note that this is a much harder setting than, e.g., images or text, where inputs have a canonical orientation. Finding the correspondence between graph nodes
 is a notoriously hard problem \cite{simonovsky2018graphvae,samanta2020nevae,you2018graphrnn,bojchevski2018netgan}, and it is exacerbated when the number of sampled graphs is large. 
To that end, we propose  AlignGraph, a group of \emph{permutation invariant} graph alignment methods combined with their application to a group of base generative models. Our main contributions are as follows:
\begin{packed_enum}
\item  \apr \  incorporates convex graph multi-distance methods in training to achieve permutation invariance. We use these tools both as means to construct alignment in a tractable fashion as well as to create soft penalties in training.
\item \apr \ is a general flexible framework. It can be applied to a broad class of base generative models, enhancing their permutation invariance. We demonstrate this here by applying it to graph recurrent neural networks~\cite{you2018graphrnn}, gated recurrent attention networks~\cite{liao2019efficient} and variational autoencoders~\cite{kipf2016variational} as our base generative models. 
\item  We propose three methods that can be parallelized to accelerate graph multi-distances. Leveraging parallelism, our methods speed up  computation by a $40\times$ factor, while maintaining the alignment accuracy and, in some cases, improving it.
\item We conduct experiments on both synthetic and real data, showing that AlignGraph outperforms both our base and other competitor models. We define two performance scores to measure accuracy. We then show that our model achieves $25\% - 250\%$ improvement in those scores over base models and $62.5\% - 4000\%$ improvement over other competitors.
\end{packed_enum} 
\section{Related Work}

\sloppy \noindent\textbf{Graph Embeddings.} Graph embeddings   map nodes  into a lower-dimensional space and have been applied to link prediction \cite{kipf2016variational,grover2016node2vec,bojchevski2018netgan}, node classification  \cite{kipf2016semi,perozzi2014deepwalk,grover2016node2vec,hamilton2017inductive}, and clustering \cite{hamilton2017inductive}. 
Many of the state-of-the-art graph embedding algorithms capture the relative position of nodes on the embedding space~\cite{gritsenko2021graph}. Because of the non-convexity of training objectives and the existence of multiple local minima, even isomorphic graphs can  map to completely different embeddings using the same embedding algorithm~\cite{gritsenko2021graph}. This is further exacerbated when graphs are near-isomorphic (i.e., differ in a few edges) as well as when the embeddings are randomized~\cite{gritsenko2021graph}. Embeddings play a central role in graph generative models (see Sec.~\ref{sec:bc:ggm}), but lack of permutation invariance can introduce significant distortions. 

\fussy 

\sloppy\noindent\textbf{Deep Generative Models.} Deep generative models can be categorized into three groups: generative adversarial networks (GANs), variational autoencoders (VAEs), and auto-regressive models. 
NetGAN~\cite{bojchevski2018netgan} 
learns the distribution of biased random walks over a single graph. GraphVAE~\cite{simonovsky2018graphvae} and NEVAE~\cite{samanta2020nevae} use VAEs linking node embeddings to edges. 
 GraphRNN~\cite{you2018graphrnn} is an auto-regressive model that constructs a graph sequentially over nodes and edges. GRAN~\cite{liao2019efficient} is another auto-regressive model that uses graph neural networks (GNNs) with an attention mechanism to generate a block of nodes and  edges  sequentially. 

\fussy Several of these methods contain techniques to partially deal with permutation invariance. For example, GraphVAE~\cite{simonovsky2018graphvae} uses an approximate graph matching to penalize misalignment between each input graph and its corresponding reconstructed graph.
NEVAE~\cite{samanta2020nevae} and GraphRNN~\cite{you2018graphrnn}
use a breadth-first-search node ordering scheme and GRAN~\cite{liao2019efficient} marginalizes over a family of canonical node orderings to handle permutation invariance. However, none of these methods address permutation invariance by finding a consistent node ordering across sampled graphs.
In comparison, the graph alignment approach we introduce here does exactly this; in addition, it is generic and can be applied to the broad group of base generative models listed above to enhance their permutation invariance (see also Sec.~\ref{sec:exp}).

\sloppy\noindent\textbf{Graph distances.}
Classic methods to compute the distance between graphs include the edit distance~\cite{garey2002computers,fischer2015approximation} and the maximum common subgraph distance~\cite{bunke1998graph,bunke1997relation}. Although they are metrics, they are hard to compute. Bento $\&$ Ioannidis~\cite{bento2018family} recently introduced a family of metrics for graph distances that is computationally tractable but limited to computing the distance between two graphs. To compute distances among  a larger group of graphs, it is important that the distance function satisfies alignment
consistency~\cite{nguyen2011optimization}. There are works on  multi-distances that enforce this constraint~\cite{huang2013consistent,chen2014near,zhou2015multi}; however, none of these methods satisfy generalizations of the metric properties. Gromov-Wasserstein
Learning (GWL), proposed by Xu et al.~\cite{xu2019gromov} satisfies both of these  properties.  However, GWL has cubic complexity and is not applicable to recurrent neural networks. Recently, two approaches were proposed by Kiss et al.~\cite{kiss2018generalization} and Safavi $\&$ Bento~\cite{safavi2019tractable} to measure the distance among a group of graphs; we describe both in detail in Sec~\ref{sec:Ga}. 
Both  satisfy alignment consistency and a generalization of metric properties~\cite{kiss2018generalization}. However, both  are also slow when applied to a large number of graphs.

\fussy We propose a framework to accelerate these two graph multi-distance algorithms, by leveraging parallelization and graph coarsening~\cite{karypis1997metis}. Graph coarsening has been used in community detection~\cite{satuluri2009scalable, dhillon2007weighted}, graph embeddings~\cite{liang2021mile} and alignment between two graphs~\cite{caper}. 
We coarsen graphs using K-means clustering, and incorporate this accelerated graph alignment method into our framework to address permutation invariance. To the best of our knowledge, we are the first to accelerate graph multi-distances by graph coarsening. 
 \section{Background} \label{sec:bc}

\subsection{Minimum Distance between two Graphs.}\label{sec:bc:SB}

Let $\mathcal{G}= (V, E)$ be an undirected graph with node set $V= [m]\equiv \{ 1, 2, \ldots, m \}$ and edge set $E \subseteq [m] \times [m]$, represented by adjacency matrix $A \in \{0, 1\}^{m \times m}$. The entries of this adjacency matrix are indexed by the nodes in $V$. We denote the set that contains all such matrices by $\Omega \subseteq \mathbb{R}^{m \times m}$. 
Consider two graphs $\mathcal{G}_A= (V, E_A)$, $\mathcal{G}_B= (V, E_B)$ with adjacency matrices $A, B \in \Omega$.
One way to measure the distance between these two graphs is to find an alignment between nodes and compute an edge discrepancy (i.e., edit distance \cite{sanfeliu1983distance,garey2002computers}) between them. An alignment  can be represented by a permutation matrix $P \in \mathcal{P}^m$, where:
\begin{align}
 \mathcal{P}^m \triangleq  \{P \in \{0,1\}^{m \times m}; \ P1=1,\ P^T1=1\}.   
\end{align}
However, finding such an alignment is generally computationally intractable \cite{bento2018family,babai2016graph}.
Bento $\&$ Ioannidis \cite{bento2018family}  introduce a distance function $d_{S}:{\Omega}^2	\longmapsto \mathbb{R} $, defined as:
\begin{align}
\label{eq:sb}
    d_{S}(A, B)= \underset{{P \in \mathcal{W}^m} }{\min} \ \|AP-PB\|+ \beta \mathrm{tr}(P^TD_{A, B}),
\end{align} 
where $\beta> 0$ is a positive regularization parameter, $\|\cdot\|$  is a matrix norm,  $\mathrm{tr}$ is the trace operator, matrix $D_{A,B} \in \mathbb{R}^{m \times m}$ represents the dissimilarity between nodes across the two graphs, and matrix $P$ is a doubly stochastic alignment matrix, that is, $P\in \mathcal{W}^m $, where
\begin{align}
\mathcal{W}^m \triangleq  \{P \in [0,1]^{m \times m}; \ P1=1,\ P^T1=1\}.
\label{eq:c}
\end{align} 
Matrix $D_{A,B}$ is generally a distance matrix, where each element represents the pairwise distances between the
embeddings or features of nodes across two graphs. For example, for two matrices of graph embeddings $Z_A \in \mathbb{R}^{m \times d}$ and $Z_B \in \mathbb{R}^{m \times d}$ that map nodes of a graph into a lower-dimensional space, i.e. $d<m$, $D_{A,B}$ is:
\begin{subequations}
\begin{align}
& D_{A, B}= [D_{a, b}]_{a \in V, b \in V} \in \mathbb{R}^{m \times m}, \ \mathrm{and}\\
 &  D_{a, b}= ||z_{a}^A-z_{b}^B||_2, \qquad \forall \ a \in V, b \in V,
  \label{eq:D}
\end{align}
\end{subequations}
where $z_{a}^A$ indicates the $a$-th row of matrix $Z_A$,  $z_{b}^B$ indicates the $b$-th row of $Z_B$. 
Intuitively, the first term in Eq.~\eqref{eq:sb}
is a probabilistic mapping between nodes of two graphs and the second term penalizes the dissimilarity between the embeddings of the nodes that are mapped to each other. Eq.~\eqref{eq:sb} is a pseudometric and a convex optimization problem \cite{bento2018family}, and thus can be computed efficiently via standard techniques.

\subsection{Minimum Distance among n Graphs.}\label{sec:Ga}
Consider a distance function $d(\mathcal{G}_i,\mathcal{G}_j)$ like Eq.~\eqref{eq:sb} that induces a (possibly stochastic) alignment matrix $P_{ij}$ between two pairs of graphs. To compute the minimum distance between a group of $n> 2$ graphs, one could simply generalize the distance function $d(\mathcal{G}_i,\mathcal{G}_j)$ to multiple graphs, via $d(\mathcal{G}_1,\mathcal{G}_2,\ldots,\mathcal{G}_n) = \sum_{i,j \in [n]} d(\mathcal{G}_i,\mathcal{G}_j)$. However, such a generalization does not guarantee the joint alignment between multiple graphs: that is, if $P_{ij}$ aligns $\mathcal{G}_i$ with $\mathcal{G}_j$, and $P_{jl}$ aligns $\mathcal{G}_j$ with $\mathcal{G}_l$, the alignment matrix $P_{il}$ should keep the consistency of alignments under transitivity, i.e., $P_{il}=  P_{ij}  P_{jl}$ \cite{safavi2019tractable}. This property is known as \emph{alignment consistency}. 
We describe next two distance functions that induce alignments that satisfy this property. 

\subsubsection{Fermat Distance.} \label{sec:bc:ferm}
Let $d(A,B)$ be a metric for two graphs such that $d: {\Omega}^2 \longmapsto \mathbb{R}$. Then the Fermat distance function \cite{kiss2018generalization} associated with $d$ is the map of $d_F: {\Omega}^n \longmapsto \mathbb{R} $ defined by:
  $  d_F(A_1,A_2,\ldots,A_n)= \underset{A_0 \in \Omega}{\min}\textstyle \sum_{i=1}^n d(A_i,A_0),$
capturing the distance among  a set of graphs. If $d$ is a metric then the Fermat distance function induced by $d$ is a so-called $n$-metric \cite{safavi2019tractable}. The 
Fermat distance function induced by Eq.~\eqref{eq:sb} is:
\begin{align}
  d_F(A_1,\ldots,A_n)=   \!\!\!\underset{\underset{{P_i \in {\mathcal{W}}^m},\forall i \in [n] }{{A_0 \in \Omega}}}{\min} \!\!\!\textstyle\sum_{i=1}^n  G(P_i,A_0; A_i ),
  \label{eq:fermat}
\end{align}
where  $D=0$ and ${\mathcal{W}}^m$ represents the set of doubly stochastic matrices and $G:{\mathcal{W}}^m \times {\Omega} \times {\Omega} \longmapsto \mathbb{R}$ is:
 \begin{align}
     G(P_i, A_0; A_i)= \|A_iP_i-P_iA_0 \|.
\label{eq:fer}     
 \end{align}
 Graph $\mathcal{G}_0$, corresponding to $A_0$, represents the center of set $\bm{\mathcal{G}}$.
The Fermat distance function in Eq.~\eqref{eq:fermat} is a pseudo $n$-metric \cite{safavi2019tractable}. This optimization problem is non-convex; nonetheless, it can be solved approximately via alternating minimization (AM): Eq.~\eqref{eq:fermat} then reduces to solving two alternating convex optimization problems. The first one has $nm^2$ parameters and $nm^2$ constraints. The second problem reduces to $n$ optimization problems with $m^2$ parameters and $m^2$ constraints. More details regarding AM iterations  can be found in \fullversion{the extended version of our paper~\cite{shayestehfard2023align}}{App.~\ref{sec: am}}. 

\subsubsection{G-align distance.} \label{sec:bc:galign}
Safavi $\&$ Bento \cite{safavi2019tractable} introduce the G-align distance function, a convex function that satisfies metric properties and alignment consistency. Consider  the map $ d_G:{\Omega}^n	\longmapsto \mathbb{R} $ defined by:


\begin{align}
     d_G({A}_1,\ldots,{A}_n)=    \underset{{P_{ij} \in S}}{\min} \ \textstyle\frac{1}{2} \sum_{i,j \in [n]} G(P_{ij};A_i, A_j) ,\!\!\!
     \label{eq:dg}
\end{align}
where $G(P_{ij};A_i, A_j)$ is given by Eq.~\eqref{eq:fer} (with $D=0$),

\begin{align}
\begin{split}
 S=\{\{P_{ij}\}_{i, j \in [n]}:  P_{ij} \in  \mathcal{P}^m, \forall i, j \in [n],\\
 P_{il}P_{lj}=P_{ij}, \forall i,j,l \in [n], P_{ii}=I, \forall i \in [n] \},
  \label{eq:s-cond}
  \end{split}
\end{align}
where $ \mathcal{P}^m$ is the set of permutation matrices and $ P_{il}P_{lj}=P_{ij}$ captures alignment consistency.

Let $\bm{P} \in R^{nm \times nm }$ be a matrix with $n^2$ blocks such that the $(i,j)$-th block is $P_{ij}$, i.e.:
\begin{align}
\label{eq:12}
 \bm{P}= \left[ \begin{smallmatrix}
    I & P_{12} & P_{13} & \dots  & P_{1n} \\
    P_{21} & I & P_{23} & \dots  & P_{2n} \\
    \vdots & \vdots & \vdots & \ddots & \vdots \\
    P_{n1} & P_{n2} & P_{n3} & \dots  & I 
\end{smallmatrix} \right].
\end{align}
Safavi $\&$ Bento~\cite{safavi2019tractable} prove that the alignment consistency is equivalent to $\bm{P} \succeq 0$ (see Lemma 4 in Safavi $\&$ Bento~\cite{safavi2019tractable}). By relaxing the permutation matrices constraint in~\eqref{eq:dg} to a doubly stochastic constraint, 
the G-align distance function is as follows:
\begin{align}
    \label{eq:18}
  d_G({A}_1,\ldots, {A}_n)=\!\!  \underset{\underset{{P_{ii}=I}, \bm{P} \succeq 0 }{{P_{ij} \in \mathcal{W}^m,}}}{\min}
  \frac{1}{2}\! \sum_{i,j \in [n]}\!\! G(P_{ij};A_i, A_j).\!\!\!
\end{align}
This is a pseudo $n$-metric (see Theorem~$5$ and Remark~$4$ in Safavi and Bento~\cite{safavi2019tractable}) and a convex optimization problem with $O(n^2m^2)$ variables and $n$ constraints. 
In practice, this problem can be solved via optimization toolboxes such as CVXPY~\cite{diamond2016cvxpy} as well as the Frank-Wolfe algorithm (FW)~\cite{frank1956algorithm}; the latter is outlined  \fullversion{in~\cite{shayestehfard2023align}.}{in App.~\ref{app:fw}.
}

Despite convexity, the quadratic nature of $\bm{P}$  (in terms of $n$ and $m$) in G-align distance and ${P_i} \in {\mathcal{W}}^m$, ${i \in [n]}$  (in terms of $m$) in Fermat distance makes these computations expensive. We address this in Section~\ref{sec:prl}. 
\subsection{Graph Generative Models.}\label{sec:bc:ggm}
Given a set of undirected graphs $\bm{\mathcal{G}}= \{\mathcal{G}_1, \mathcal{G}_2, \ldots, \mathcal{G}_n \}$ sampled from $p(G)$, for each graph $\mathcal{G}_i(V_i, E_i)$, $\forall i \in [n]$, we denote the adjacency matrices by $A_i \in \mathbb{R}^{m \times m}$ and  feature matrices by $X_i\in \mathbb{R}^{m \times f}$. Features could be either one-hot node indicator vectors or consist of graph characteristics, such as, e.g., node degrees. The nodes of each graph are mapped to a latent embedding space via a deep neural network, parameterized by $\phi$~\cite{hamilton2017inductive}: 
\begin{align}
    Z_i= f_{\phi}(A_i, X_i),
    \label{eq:enc_param}
\end{align}
where $Z_i= \{z_{i_1}, z_{i_2}, \cdots, z_{i_m} \}$ denotes the  hidden node representations and $\phi$ represents parameters of the deep neural network encoder. The  decoder parameterized by $\theta$ takes the hidden representations and reconstructs the adjacency matrix, i.e.:
\begin{align}
    \hat{A}_i = g_{\theta}(Z_i),
    \label{eq:dec_param}
\end{align}
where $\hat{A}_i$ is the estimated adjacency matrix. Both the encoder and decoder can be randomized, and induce a distribution $p_{\phi,\theta}$ over graphs. A  loss often used  to train parameters over graphs is the negative log likelihood: \begin{align}
    L(\phi, \theta; \bm{A}, \bm{X})= -\textstyle\sum_{i=1}^n \mathrm{log}\, p_{\phi,\theta}(A_i), \label{eq:gen_loss}
\end{align}
where $\bm{A}= \{A_1, A_2, \ldots, A_n\}$ denotes the set of adjacency matrices and $\bm{X}= \{X_1, X_2, \ldots, X_n\}$ represents the set of feature matrices.

Several existing generative models can be described using the general framework described by Eq.~\eqref{eq:enc_param}-\eqref{eq:gen_loss}.
GraphRNN~\cite{you2018graphrnn} is an auto-regressive model that generates node embeddings sequentially: $f_{\phi}$~\eqref{eq:enc_param} is an RNN that encodes the states of graph generated so far, and $g_{\theta}$~\eqref{eq:dec_param} is a Gated Recurrent
Unit (GRU) model that outputs the distribution of the next node's adjacency vector. 
GRAN~\cite{liao2019efficient} is also an auto-regressive model that generates the graph in a block by block basis. In this model, $f_{\phi}$~\eqref{eq:enc_param} is a GRU that uses an attention-weighted sum over the neighborhood of each node to produce the corresponding node embedding and $g_{\theta}$~\eqref{eq:dec_param} models the probability of generating edges in a block comprising multiple rows of a graph adjacency matrix via a mixture of Bernoulli distributions. 
In VAE~\cite{kipf2016variational}, $f_{\phi}$~\eqref{eq:enc_param} is a probabilistic encoding denoted by $q_{{\phi}}(Z_i| A_i, X_i)$. There is a prior over the latent variables $p_z(Z_i) \sim N(0, I)$ and $g_{\theta}$~\eqref{eq:dec_param} is defined as the inner product between latent variables.
The loss function~\eqref{eq:gen_loss} is further approximated by a variational lower bound of the log-likelihood \cite{kingma2013auto}.
All these models can be trained by minimizing the loss~\eqref{eq:gen_loss} via standard gradient methods.

\sloppy



\fussy
\section{AlignGraph}\label{sec:method}
We present \apr, our framework for enhancing the permutation invariance of base generative models. \apr can be applied to any base generative model of the form given by Eq.~\eqref{eq:enc_param}- \eqref{eq:gen_loss}: we indeed apply it to GraphRNN~\cite{you2018graphrnn}, GRAN~\cite{liao2019efficient} and VAE~\cite{kipf2016variational}  in Sec.~\ref{sec:exp}.
We consider three \apr variants, described next.

\subsection{\aprone.} \label{sec:gsv}
We begin by aligning sampled graphs. To do this, we first compute $\bm{P}$ by solving the problem in Eq.~\eqref{eq:18} \cite{safavi2019tractable}.
We  take the first block column of $\bm{P}$, i.e., $\{P_{i1}\}_{i=1}^n$, and project each $P_{i1} \in \mathcal{W}^m$ onto the set of permutation matrices, i.e.: 
\begin{align}
 \tilde{P_{i1}}= {\Pi}_{\mathcal{P}^m} (P_{i1}),   
 \label{eq:hun}
\end{align}
where ${\Pi}_{\mathcal{P}^m}$ is the orthogonal projection to $\mathcal{P}^m$. This can be done in polynomial time with the Hungarian algorithm \cite{kuhn1955hungarian}. Given these permutation matrices, we align all graphs and features with the first graph, via:
\begin{align}
 \tilde{A_i}= \tilde{P_{i1}}^TA_i\tilde{P_{i1}}, \quad \tilde{X_i}= \tilde{P_{i1}}^TX_i \quad \forall i \in [n].
 \label{eq:align-all}
  \end{align} 
Note that, by alignment consistency ~\eqref{eq:s-cond}, this could be done on any block column; our selection of $\{P_{i1}\}_{i=1}^n$ is arbitrary. Given the alignment matrices $\{\tilde{P_{i1}}\}_{i \in [n]}$, the adjacency matrices $\textbf{A}$ and feature matrices $\textbf{X}$, we aim to solve the following optimization problem: 
  \begin{align}
 \underset{\phi, \theta}{ \min}  \textstyle \frac{1}{n}\sum_{i=1}^n L({\phi}, {\theta}; \tilde{A_i}, \tilde{X_i}),
 \label{eq:L3-esp-opt}
\end{align}
where $\phi$ and $\theta$ are the DNN parameters and $L(\cdot)$ is the loss defined in Eq.~\eqref{eq:enc_param}-\eqref{eq:gen_loss}. 
\sloppy Eq.~\eqref{eq:L3-esp-opt} can be solved via stochastic gradient descent (SGD). 

\subsection{\aprtwo.} \label{sec:gdv}
In our second approach, we (a) compute a central graph across the graph set and (b)  enforce that this graph and aligned graphs are jointly   embeddable in the same space. To that end, we use two base generative models combined with the G-align distance function. We again compute $\bm{P}$ from the G-align distance~\cite{safavi2019tractable}  by solving Eq.~\eqref{eq:18}  and project each $P_{i1} \in \mathcal{W}^m$ onto the set of permutation matrices $\{\tilde{P_{i1}}\}_{i \in n}$ via \eqref{eq:hun}. We then align the adjacency matrices and feature matrices as in Eq.~\eqref{eq:align-all}.
We then estimate the center graph $\mathcal{G}_0$ of set $\bm{\mathcal{G}}$, i.e. the graph which has the minimum distance from all the graphs in the graph set via:
\begin{align}
      \underset{  {\hat{A}_{0_{j,k}, }\in [0,1], \forall j , k \in [m]} }\min  \textstyle\sum_{i=1}^n \|\tilde{A_i}- \hat{A_0}\|.
      \label{eq: a0-2}
\end{align}
Prob.~\eqref{eq: a0-2} is  convex  and can be solved via standard methods: the objective has $n$ terms, and the problem has $m^2$ parameters and $O(m^2)$ constraints. Since $\hat{A}_{0_{j,k}}\in [0,1], \ \forall j,k \in [m] $, we binarize the elements of the adjacency matrix for $\mathcal{G}_0$ by using a threshold. Once we estimate $\mathcal{G}_0$, given the permutation matrices $\{\tilde{P_{i1}}\}_{i \in n}$ and the graph set $\bm{\mathcal{G}}$ with one-hot encoding feature matrices $\textbf{X}$, we train two  generative models of the chosen type. We train the first with $\mathcal{G}_0$  and  the second with all aligned $\{\mathcal{G}_i\}_{i=1}^n$ . We train the generative models jointly, by penalizing the distance between the embeddings of $\mathcal{G}_i$ and $\mathcal{G}_0$, i.e.:

\begin{align}
\begin{split}
    \underset{ \Phi, \Theta}{\min}  & \quad
     \textstyle\frac{1}{n} \sum_{i=1}^n [L({\phi}, {\theta}; \tilde{A_i}, \tilde{X_i})+\beta \trace( D(\tilde{Z_i}, Z_0))] \\
   &  + L({\phi}_0, {\theta}_0, A_0, X_0) ,
     \end{split}
    \label{eq:L6}
\end{align}
where $\beta>0$ is a positive regularization parameter,  $L(\cdot)$ is a loss function of a base generative model defined in Eq.~\eqref{eq:enc_param}-\eqref{eq:gen_loss}, $\Phi=\{ \phi_0, \phi \}$ and $\Theta= \{\theta_0,\theta \}$ are the generative models parameters,  $Z_0 , \{Z_i\}_{i \in [n]} \in \mathbb{R}^{m \times d}$ are the hidden representation of nodes and $D$ is given by Eq.~\eqref{eq:D}. Note that the trace enforces the joint embeddability of all graphs with the central graph. The objective in Eq.~\eqref{eq:L6} again can be minimized via SGD. 
After training we take \emph{only} the generative model parameterized by $\phi, \theta$ to generate new graphs.

\subsection{\aprthree.} \label{sec:fdv}
In this model, we combine the Fermat distance function with two similar-structure generative models. We first use the Fermat distance function defined in Eq.~\eqref{eq:fermat} to estimate graph alignment matrices  $\{P_{i}\}_{i \in [n]}$ and $\mathcal{G}_0$ via alternating minimization. Then, we project each $P_{i} \in \mathcal{W}^m$ onto the set of permutation matrices $\{\tilde{P_{i}}\}_{i \in n}$ via Eq.~\eqref{eq:hun}. Given the graph set $\bm{\mathcal{G}}$, the center graph $\mathcal{G}_0$ and alignment matrices $\{\tilde{P}_{i}\}_{i \in [n]}$, we train the two generative models jointly. We train the first with  $\mathcal{G}_0$ and the second with the aligned $\{\mathcal{G}_i\}_{i \in [n]}$. We minimize the distance between the embeddings of these two generative models by solving the following optimization problem:
\begin{align}
 \begin{split}   \underset{ \Phi, \Theta}{\min} & \quad
  \textstyle \frac{1}{n}\sum_{i=1}^n
    [L({\phi}, {\theta}; \tilde{A_i}, \tilde{X_i})+ \beta \trace(D(\tilde{Z_i}, Z_0))]\\
    & +L({\phi}_0, {\theta}_0, A_0, X_0),\end{split}\label{eq:l4}
\end{align}
where $\beta>0$ is a positive regularization parameter, $L(\cdot)$ is again a loss function of a base generative model defined in Eq.~\eqref{eq:enc_param}-\eqref{eq:gen_loss}, $\Phi=\{\phi_0,\phi \}$ and $\Theta= \{\theta_0,\theta \}$ are the generative models parameters, $Z_0, \{Z_i\}_{i \in [n]} \in \mathbb{R}^{m \times d}$ are graph embeddings and $D$ is given in Eq.~\eqref{eq:D}.
We again solve Eq.~\eqref{eq:l4} w.r.t.~$\Phi$ and $\Theta$ via SGD. After training, we  again use only the generative model parameterized by $\phi$ and $\theta$ to generate graphs.

\subsection{Extensions.} Our proposed graph alignment methods are not limited to graphs with equal numbers of nodes; they can be readily extended to collections of graphs with a variable number of nodes by employing one of several ways to add ``dummy'' nodes such that all graphs have equal number of nodes~\cite{bento2018family}. A simple solution is to first find the maximum number of nodes $\mathrm{m}_\mathrm{max}$ in the graph set and then expand all graphs with $|V_i| <\mathrm{m}_\mathrm{max}$, $ i \in [n]$ by adding ``dummy'' nodes such that all graphs have $\mathrm{m}_\mathrm{max}$ nodes. In the expanded graphs ``dummy'' nodes are connected to each other as well the actual nodes by edges with a small weight (e.g., $0.01$) to differentiate these edges from the edges connecting the actual nodes.


\section{Accelerated  Multi-Distances.} \label{sec:prl}
In both Fermat distance and G-align distance, as the number $n$ of graphs grows, alignment becomes more computationally  expensive. 
We propose three methods to accelerate multi-distance algorithms. All methods produce a  final center graph, $\mathcal{G}_{0_{\mathrm{out}}}$; once this is computed, all  the graphs in $\bm{\mathcal{G}}$ can be aligned with $\mathcal{G}_{0_{\mathrm{out}}}$ (and each other) via Eq.~\eqref{eq:sb}. We describe these methods assuming alignment happens via the G-align distance, but the methods extend, mutatis mutandis, to Fermat distance as well, by replacing Eq.~\ref{eq:18} with Eq.~\ref{eq:fermat}. We provide pseudocode for all three methods  \fullversion{in~\cite{shayestehfard2023align}.}{in App.~\ref{app:acmd}.}

\noindent\textbf{G-Parallel: Grouping and Parallelizing Graphs.} This method has a recursive structure, comprising $O({\log}_{K} n)$ stages, where $K\in \mathbb{N}$. In each  stage, we apply the same three-step procedure on a smaller set of graphs, starting from the full set of graphs in the training set. In the first step, we  divide the set of graphs  into a collection of  smaller groupings of size $K\ll n$. In the second step, we compute the alignment via Eq.~\eqref{eq:18} \emph{within each group}. In the third step, we output a center graph, computed via Eq.~\eqref{eq: a0-2}, for each group. Note that the operations in the second and third steps can happen in parallel. The procedure then executes recursively  on the (smaller) set of center graphs. The output of the final stage is a single center graph, $\mathcal{G}_{0_{\mathrm{out}}}$.  We note that, for  Eq.~\eqref{eq:fermat}, Eq.~\eqref{eq:18}, and Eq.~\eqref{eq: a0-2}, computing alignments over $K\ll n$ rather than $n$ graphs yields significant performance dividends even serially, because the execution cost is super-quadratic in the number of graphs. The total number of such $K$-graph problems  we compute is $O(\frac{n}{K})$.

\noindent \textbf{C-Serial: Coarsening Graphs.} 
In this method, we create coarsened graphs~\cite{karypis1997metis} by partitioning each graph into $c \in \mathbb{N}$ clusters via clustering algorithm such as K-means. In short, the nodes in a coarsened graph are super-nodes representing all nodes in the original graphs' clusters. The weighted edges are the unions of edges connecting two clusters in the original graph. We next compute the graph alignment across the \emph{coarsened graphs}, via Eq.~\eqref{eq:18}. 
Having mapped clusters to each other across graphs, we refine alignments: we align the nodes within the clusters via Eq.~\eqref{eq:18} on a per-cluster basis. This yields a global alignment;
finally, we construct a center graph by computing the center for the clusters and the edges connecting the clusters via Eq.~\eqref{eq: a0-2}.  In this method, we need to compute distances over $O(n)$ graphs again but of size $O(c)$, with the refinement involving $nc$ pairwise alignments of size, approximately, $m/c$, assuming clusters of equal size. 

 \noindent\textbf{CG-Parallel: Coarsening, Grouping and Parallelizing.} Similar to G-Parallel, this method is recursive and in each stage we apply the same procedure on a smaller set of graph. We just change what happens in each stage compared to G-Parallel. 
 Again, similar to G-Parallel, in each stage we first divide graphs into smaller groupings. In each of these smaller groups, we compute the center graphs exactly the same way we did in C-serial, i.e., by coarsening graphs, computing the alignments via Eq.~\eqref{eq:18}, computing the center graph by computing the center of clusters  and edges connecting clusters via Eq.~\eqref{eq: a0-2}. The procedure then executes
recursively on the (smaller) set of center graphs.
 The output of the final stage is a center graph, $\mathcal{G}_{0_{\mathrm{out}}}$, for the whole set. The total number of stages in this method is $O({\log}_{K} n)$. The total number of such K-graph problems we
compute is $O(\frac{n}{K})$ with the refinement involving $Kc$ pairwise alignments of size, approximately, $m/c$, assuming clusters of equal size.
 

\section{Experimental Setup} \label{sec:exp}

\subsection{Datasets.}
We perform experiments on both synthetic and real datasets with varying numbers of nodes and edges, using the code in~\cite{you2018graphrnn}. 

\begin{table}[!t]
    \begin{scriptsize}
    \begin{tabular}{*{5}{c}}
        \toprule
        & $|V|_{\mathrm{ave}}$ & $|E|_{\mathrm{ave}}$ & $n$ & Alignment alg. \\
        \midrule
Community (small)& $45$ & $98$ & $100$ & G-Parallel  \\ Community (large)&  $150$&  $2727$& $100$ &  CG-Parallel \\
Grid& $36$& $265$& $100$ & G-Parallel  \\
Ego-Citeseer & $35$& $65$ & $100$  & G-Parallel\\ 
Ego-B-A (small) & $118$& $298$& $100$  & CG-Parallel \\ 
Ego-B-A (large)& $1028$& $1471$ & $68$& CG-Parallel \\ 
Protein & $117$  &$280$& $100$ & CG-Parallel\\
\bottomrule
\end{tabular}
\end{scriptsize}
\caption{Dataset summary including average number of nodes and edges and number of graphs in the graph set, along with the algorithm used to compute graph alignment. For smaller graphs (with $|V|_{\mathrm{ave}} < 50$ ) we use the G-Parallel method. For larger graphs, to further accelerate computing the graph alignment, we use CG-Parallel. For all parallel alignment algorithms we use a single machine with $40$ CPUs.
}
\label{tab:Datasets}
\end{table}

\noindent\textbf{Community.}
We generate two community graphs, with three-communities from the stochastic block model~\cite{you2018graphrnn}. The first graph has ${|V|}= 45$ total nodes and $[5, 15, 17]$ nodes in the communities. The second has ${|V|}= 150$ total nodes and $[40, 50, 60]$ nodes in the communities. In both graphs, each community is generated by the Erd\H{o}s-R\'{e}nyi model (E-R)~\cite{erd1959and}. The probability for edge creation in each community is $p= 0.7$. For the smaller graph $0.05 |V|$ inter-community edges were added and for the large community graph $0.005 |V|$ inter-community edges were added u.a.r. 
In order to build the graph set, we generate $100$ random graphs by randomly permuting the graph and then add noise by randomly removing and re-adding $10\%$ of edges, selected u.a.r.

\noindent\textbf{Grid.}
We construct a $2$-$\mathrm{D}$ grid graph with $|V|= 36$ nodes. As above, we generate $100$ graphs by randomly permuting the graph and again add noise by randomly removing and re-adding $10\%$ of edges, u.a.r.

\noindent\textbf{Ego-B-A (small).} We generate $100$ graphs with $|V|= 950$ nodes using the Barab\'{a}si-Albert model. During the generation of each graph, each node in a graph is connected to $5$ existing nodes.  We then construct $1-$hop ego graphs with $|V| \in [100-130]$ nodes.

\noindent\textbf{Ego-B-A (large)}. We generate $68$ graphs using the Barab\'{a}si-Albert model. Each graph has $|V|= 75500$ nodes such that each node is connected to $5$ existing nodes during generation. In the next step, we construct $1-$hop ego graphs with $|V| \in [1000-1050]$ nodes.

\noindent\textbf{Ego-Citeseer.} Similar to ~\cite{you2018graphrnn, tran2020deepnc}, we construct $100$ $3$-hop ego graphs from the Citeseer network~\cite{sen2008collective}, with $|V| \in [30-40]$ nodes.  

\noindent\textbf{Protein.}  Similar to ~\cite{you2018graphrnn, liao2019efficient}, we select $100$ protein graphs from a protein dataset~\cite{dobson2003distinguishing} with $|V| \in [100, 130]$ nodes. The nodes in these graphs represent amino acids and the edges are placed between  all pairs of nodes  that are less than 6 Angstroms apart.

Table~\ref{tab:Datasets} summarizes each dataset as well as  graph set size and the methods used to compute graph alignments. In all datasets, we use CVXPY~\cite{diamond2016cvxpy} as our solver; additional implementation details can be found \fullversion{in~\cite{shayestehfard2023align}}{in App.~\ref{app:acmd}}.

\subsection{Algorithms.}
We compare our methods against three base generative models, GraphRNN~\cite{you2018graphrnn}, GRAN~\cite{liao2019efficient}, 
and VAE~\cite{kipf2016variational} and two competitors, GraphVAE~\cite{simonovsky2018graphvae} and DeepGMG~\cite{li2018learning}.  Additional details on baseline algorithms are in \fullversion{\cite{shayestehfard2023align}}{App.~\ref{app:impd}}.
We compare these baselines to all three versions of \apr \ described in Sec.~\ref{sec:method}, where for each of our algorithms we test with three base generative models (GraphRNN~\cite{you2018graphrnn}, GRAN~\cite{liao2019efficient}, VAE~\cite{kipf2016variational}). Our code is publicly available.\footnote{
\url{https://github.com/neu-spiral/AlignGraph}}

\subsection{Performance Metrics}. \label{metrics}
In all experiments we take $80\%$ of the full set of graphs for training and use the rest for testing. We train our generative models on the training set, and use them to generate a set of synthetic graphs, whose  properties we then compare to graphs in the test set to evaluate whether the generated graphs are likely to have come from the same distribution as the test set.
We use two performance metrics to assess the quality of the generated graphs.  In both metrics, we first calculate a set of summary statistics from each individual graph (e.g., degree distribution, clustering coefficient, etc.); we summarise these statistics in \fullversion{\cite{shayestehfard2023align}}{App.~\ref{app:met}}. 
Then we compare the distributions of these statistics between the generated and test graphs w.r.t.~two metrics. The first is
the {$\mathrm{s}_{\mathrm{mmd}}$ score:} this score, proposed by You et al.~\cite{you2018graphrnn}, measures the maximum mean discrepency (MMD) between two distributions of graph statistics. The $s_{\mathrm{mmd}}$ takes values in $[0, 1]$ (the  smaller the better). We calculate an average MMD across all the statistics; a formal definition can be found   \fullversion{in~\cite{shayestehfard2023align}}{in App.~\ref{perf}}.

The second performance score is the {$\mathrm{s}_{\mathrm{mvr}}$ score:} this measures the squared difference between the mean values of the two distributions, rescaled by the variance of the value over the ground truth graphs. 
This score takes values in $[0, \infty]$  (the smaller the better). Again, we average this across all statistics (see also~\fullversion{\cite{shayestehfard2023align}}{App.~\ref{perf}}).

We also report the time it took to compute graph alignments, $t_a$, and the total training time of generative models, $t_{\mathrm{tr}}$. We compute graph alignment only once and pre-align graphs before training our \apr models. We measure $t_a$ and $t_{\mathrm{tr}}$ to have a fair comparison between the improvement we might get in $\mathrm{s}_{\mathrm{mmd}}$ and $\mathrm{s}_{\mathrm{mvr}}$ and the cost of this improvement in terms of the total time consumed by each model.

To evaluate the performance of our  accelerated multi-distances, we measure the accuracy of alignments. For this purpose, we first compute graph alignment and the center graph via Eq.~\eqref{eq:fermat} for Fermat distance and Eq.~\eqref{eq:18} and Eq.~\eqref{eq: a0-2} for G-align distance. We then align the graphs in the graph set w.r.t $\mathcal{G}_0$ and evaluate the distance of $\mathcal{G}_0$ from the graph set via
 $d_0= \frac{1}{n} \sum_{i=1}^n \textstyle \frac{\|P_i^TA_iP_i-A_0\|}{\|A_1\|}$
  \label{eq:d0}
(smaller is better).

\subsection{Results} \label{sec:exp:results}\hfill\\

\noindent\textbf{Accelerated Multi-distances Speed and Accuracy.}
 We investigate the impact of our methods on running time and on the accuracy of graph alignment computation 
on a graph set with $12$ $3-$community graphs of $|V|= 45$ nodes.  
\begin{figure}[!t]
    \centering
    \begin{subfigure}[t]{0.49\linewidth}
        \centering
        \includegraphics[width=.99\linewidth]{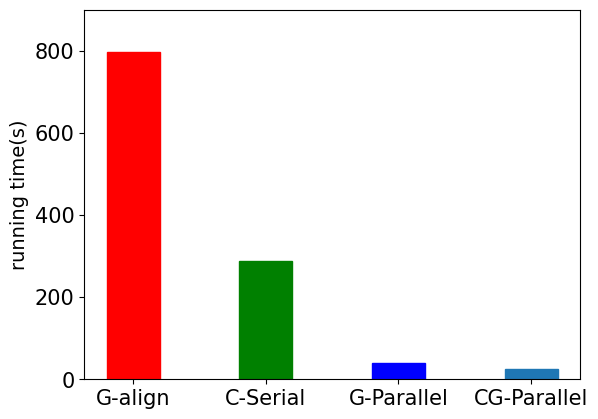}
        \caption{Running time (in seconds) using G-align distance and applying our proposed methods to G-align distance.}
        \label{fig:time1}
    \end{subfigure}
    \hfil
    \begin{subfigure}[t]{0.49\linewidth}
        \centering
        \includegraphics[width=.99\linewidth]{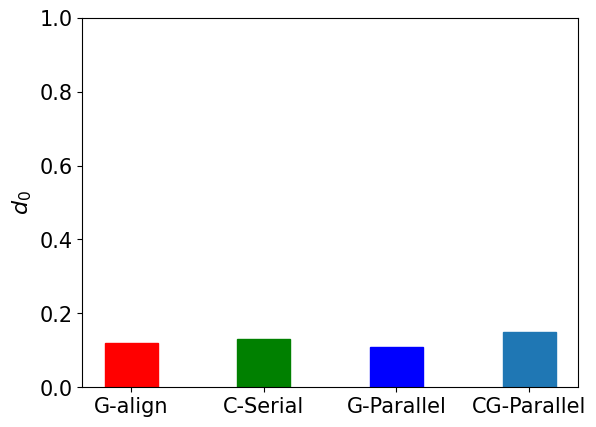}
        \caption{Accuracy of G-align distance and applying our proposed accelerated multi-distances to Fermat distance.}
        \label{fig:acc1}
    \end{subfigure}

    \begin{subfigure}[t]{0.49\linewidth}
        \centering
        \includegraphics[width=.99\linewidth]{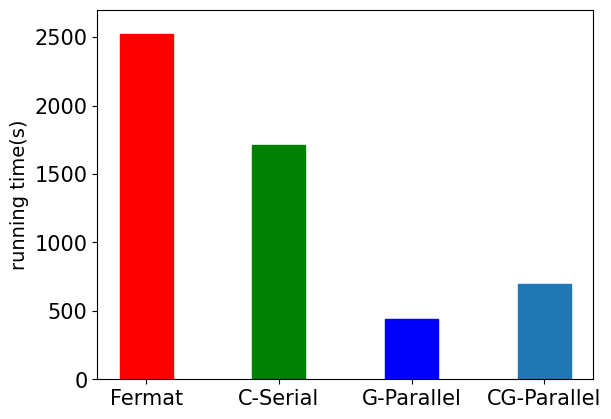}
        \caption{Same as part (a) but for Fermat distance.}
        \label{fig:time2}
    \end{subfigure}
    \hfil
    \begin{subfigure}[t]{0.49\linewidth}
        \centering
        \includegraphics[width=.99\linewidth]{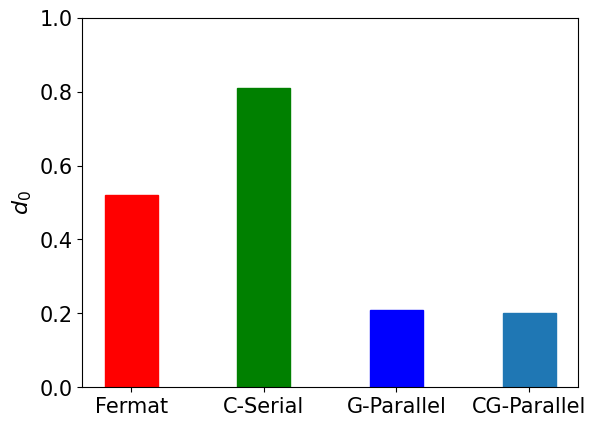}
        \caption{Same as part (b) but for Fermat distance.}
        \label{fig:acc2}
    \end{subfigure}
    
    \caption{Computation time and accuracy of computing the graph alignment in community graphs given the baselines and our three accelerated multi-distances, G-Parallel (40 CPUs), CG-Parallel (40 CPUs), and C-Serial. G-align distance has better performance compared to Fermat distance. Moreover, due to the clustered structure of community graphs  clustering and grouping graphs in CG-Parallel also improves the accuracy.}
    \label{fig:Naive}
\end{figure}
In Fig.~\ref{fig:time1} and Fig.~\ref{fig:time2} we report the total time to compute the alignment using G-align distance and Fermat distance, respectively. 
These figures demonstrate that our proposed methods reduces the computation time by $40$ times. In  Fig.~\ref{fig:acc1} and Fig.~\ref{fig:acc2} we compute $d_0$ via Eq.~\ref{eq:d0}. Our results illustrate that our acceleration methods improve the accuracy of estimated center graphs. Since G-Parallel and CG-Parallel have the best trade offs for the running time and accuracy, we use these two methods to compute graph alignment in our next experiments.

\noindent\textbf{Evaluating the Generated Graphs.} Table \ref{table:4} summarizes the performance scores $s_{\mathrm{mmd}}$ and $s_{\mathrm{mvr}}$ on all $7$ datasets. Our experiments show that our model achieves $25\% - 250\%$ accuracy improvement over base models and $62.5\% - 4000\%$ improvement over other competitors. In some datasets, such as Community graphs and Protein graphs, \aprtwo and \aprthree that jointly train two similar structure generative models produce the best performance scores. In the majority of the experiments, applying our frameworks to either base GraphRNN or base GRAN leads to the best performance scores. However, there is no clear winner between these two base generative models.
Our results in Table \ref{table:4} illustrate that our accelerated multi-distances methods scale well to larger graphs and are compatible with large datasets with $|V| > 1000$. Moreover, comparing the $t_{\mathrm{tr}}$ of \aprone (GraphRNN) and \aprone (GRAN) models with their baselines demonstrate that our models are $4.21\%-44\%$ faster. This happens due to the pre-alignment of graphs in our models. On the other hand, the ${t_a}/{t_{\mathrm{rm}}}$ ratio for  \aprone (GraphRNN) and \aprone (GRAN) models  ranges from $0.89\%$ to $150\%$, where $150\%$ belongs to the alignment of our largest dataset, Ego-B-A (large). While this pre-alignment took $70$ minutes, it led to at least $83\%$ improvement in the performance scores.

\begin{table*}[!t]
    \centering
    \setlength{\tabcolsep}{2pt}
    \resizebox{\textwidth}{!}{
    \begin{tabular}{l c c c c c c c c c c c c c c c c c c c c c c c c c c c c c c c c c c c c c c}
         \toprule
        \multirow{5}{*}{} &
        \multicolumn{9}{c}{\emph{ $\mathrm{Community \ Graphs}$}} & \multicolumn{1}{c}{} & \multicolumn{4}{c}{\emph{ $\mathrm{Grid  \ Graphs}$}} & \multicolumn{1}{c}{} & \multicolumn{4}{c}{\emph{ $\mathrm{Ego-Citeseer  \ Graphs}$}} & \multicolumn{1}{c}{} &
        \multicolumn{9}{c}{\emph{ $\mathrm{Ego-B-A  \ Graphs}$}} & \multicolumn{1}{c}{} & \multicolumn{4}{c}{\emph{ $\mathrm{Protein  \ Graphs}$}}
        \\ \cmidrule{2-10} \cmidrule{12-15} \cmidrule{17-20} \cmidrule{22-30} \cmidrule{32-35}
        \multicolumn{1}{c}{}  &
        \multicolumn{4}{c}{\emph{$(|V|_{\mathrm{ave}}$, $|E|_{\mathrm{ave}})$}}  &
        \multicolumn{1}{c}{}  &
        \multicolumn{4}{c}{\emph{$(|V|_{\mathrm{ave}}$, $|E|_{\mathrm{ave}})$}}&
         \multicolumn{1}{c}{}  &
        \multicolumn{4}{c}{\emph{$(|V|_{\mathrm{ave}}$, $|E|_{\mathrm{ave}})$}}  &
        \multicolumn{1}{c}{}  &
        \multicolumn{4}{c}{\emph{($|V|_{\mathrm{ave}}$, $|E|_{\mathrm{ave}})$}} &
        \multicolumn{1}{c}{}  &
        \multicolumn{4}{c}{\emph{$(|V|_{\mathrm{ave}}$, $|E|_{\mathrm{ave}})$}}  &
        \multicolumn{1}{c}{}  &
        \multicolumn{4}{c}{\emph{$(|V|_{\mathrm{ave}}$, $|E|_{\mathrm{ave}})$}} 
        &
        \multicolumn{1}{c}{}  &
        \multicolumn{4}{c}{\emph{$(|V|_{\mathrm{ave}}$, $|E|_{\mathrm{ave}})$}}
        \\ 

        \multicolumn{1}{c}{}  &
        \multicolumn{4}{c}{\emph{$(45, 98)$}}  &
        \multicolumn{1}{c}{}  &
        \multicolumn{4}{c}{\emph{$(150, 2727)$}}&
         \multicolumn{1}{c}{}  &
        \multicolumn{4}{c}{\emph{$(36, 265)$}}  &
        \multicolumn{1}{c}{}  &
        \multicolumn{4}{c}{\emph{$(35, 65)$}} &
        \multicolumn{1}{c}{}  &
        \multicolumn{4}{c}{\emph{$(118, 298)$}}  &
        \multicolumn{1}{c}{}  &
        \multicolumn{4}{c}{\emph{$(1028, 1471)$}}  &
        \multicolumn{1}{c}{}  &
        \multicolumn{4}{c}{\emph{$(117,  280)$}}
        \\ \cmidrule{2-10} \cmidrule{12-15} \cmidrule{17-20} \cmidrule{22-30} \cmidrule{32-35}
        
         & \emph{$\mathrm{s}_{\mathrm{mmd}}$} & \emph{$\mathrm{s}_{\mathrm{mvr}}$} & $\mathrm{t_{tr} (min)}$ & $\mathrm{t_{a} (min)}$ & & \emph{$\mathrm{s}_{\mathrm{mmd}}$} & \emph{$\mathrm{s}_{\mathrm{mvr}}$} &$\mathrm{t_{tr} (min)}$ & $\mathrm{t_{a} (min)}$ &  & \emph{$\mathrm{s}_{\mathrm{mmd}}$} & \emph{$\mathrm{s}_{\mathrm{mvr}}$} &$\mathrm{t_{tr} (min)}$ & $\mathrm{t_{a} (min)}$ &  & \emph{$\mathrm{s}_{\mathrm{mmd}}$} & \emph{$\mathrm{s}_{\mathrm{mvr}}$}&$\mathrm{t_{tr} (min)}$ & $\mathrm{t_{a} (min)}$ &   & \emph{$\mathrm{s}_{\mathrm{mmd}}$} & \emph{$\mathrm{s}_{\mathrm{mvr}}$}&$\mathrm{t_{tr} (min)}$ & $\mathrm{t_{a} (min)}$ &  & \emph{$\mathrm{s}_{\mathrm{mmd}}$} & \emph{$\mathrm{s}_{\mathrm{mvr}}$}& $\mathrm{t_{tr} (min)}$& $\mathrm{t_{a} (min)}$ 
         &  &
         \emph{$\mathrm{s}_{\mathrm{mmd}}$} & \emph{$\mathrm{s}_{\mathrm{mvr}}$}& $\mathrm{t_{tr} (min)}$ & $\mathrm{t_{a} (min)}$
         \\

        \midrule
        

       \rowcolor{anti-flashwhite} GraphVAE &$0.20$ & $612.17$& $5229.60$ & $0$& &$-$ & $-$& $-$ & $-$ & &$0.13$& $13.39$& $3827.79$ &$0$& & $0.04$ &$0.66$ & $4084.49$ &$0$&  &$-$& $-$& $-$ &$-$& & $-$ &$-$ & $-$& $-$& & $-$& $-$& $-$&$-$\\[0.1cm]
       \rowcolor{almond} DeepGMG & $0.15$ & $1180.66$& $2771.26$ &$0$&  & $-$ & $-$ & $-$& $-$&  & $0.18$  & $7.16$ & $2771.41$ &$0$& & $0.01$ &$0.92$ & $2771.47$ &$0$&  & $-$ & $-$ & $-$ &$-$& & $-$ & $-$ & $-$ &$-$& & $-$ &$-$ &$-$& $-$ \\
    \rowcolor{bisque} VAE & $0.18$ & $895.42$ & $0.13$ &$0$ & & $0.22$ & $6475.40$ & $0.86$ & $0$&  & $0.24$ & $66.09$ & $0.06$ &$0$&  & $0.06$ & $24.22$ & $0.07$ &$0$& & $0.16$ & $375.55$ & $0.18$& $0$& & $0.25$ & $22110.65$& $26.64$ & $0$& & $0.12$& $8.05$  & $19.40$& $0$\\[0.1cm]
    
       \rowcolor{burlywood}  GraphRNN & $0.08$ & $50.11$ & $45.30$ & $0$& & $0.14$  & $1270.27$  & $213.16$ &$0$& & $0.10$ & $11.21$  &$45.73$ &$0$& & $0.005$ & $0.19$ & $52.45$ &$0$& & $0.018$ & $2.78$ & $75.86$ &$0$& &$-$ & $-$ &$-$ &$-$&  & $0.06$  & $1.28$  & $75.19$ & $0$ \\[0.1cm]

        \rowcolor{vanilla} GRAN & $0.017$ & $26.98$ & $77.33$ &$0$& & $0.16$ & $4611223.41$ & $525.76$ &$0$& & $0.12$  & $27.64$ & $76.69$&$0$& & $0.009$ & $0.70$ & $76.17$ &$0$& &$0.011$  & $0.60$  & $19.54$ &$0$& &$0.011$ & $31.26$ &$50.44$ &$0$& & $0.07$ & $1.61$ & $22.03$& $0$ \\[0.1cm]        
        
        \midrule 
        
    \rowcolor{bisque} \aprone (VAE) & $0.16$  & $357.20$ & $9.72$ &$25.5$& & $0.19$ & $4983.63$  & $72.70$ & $50.41$ & & $0.20$  & $12.52$ & $5.04$ & $5.08$& & $0.02$ & $5.37$ & $8.32$ & $8.86$& & $0.10$ & $63.56$  & $16.76$ & $2.99$& &$0.24$ &$22131.28$ &$70.45$ &$70.36$ & & $0.15$ & $4.47$ & $22.31$& $4.7$ \\ [0.1cm]
     \rowcolor{bisque}  \aprtwo (VAE) & $0.09$  & $\textbf{8.76}$  & $9.12$ &$25.5$& & $0.16$  & $488.93$ & $126.35$ &$50.41$ & & $0.17$ & $39.72$  & $6.97$ &$5.08$& & $0.04$  & $2.81$ & $8.41$ & $8.86$& & $0.03$& $38.08$ & $108.29$ &$2.99$& &$-$ &$-$ &$-$ & $-$ & & $\textbf{0.03}$ & $\textbf{0.90}$ & $137.26$ & $4.7$\\ [0.1cm]
     \rowcolor{bisque}   \aprthree (VAE) & $0.09$  & $1543.08$ & $8.84$ & $6.33$& & $0.18$ & $138.39$  & $77.90$& $77.90$& & $0.18$  & $45.19$  & $5.70$ & $3.18$& & $0.02$  & $13.37$  & $7.21$ & $5.69$& & $0.07$ & $380.99$ & $95.44$& $125.4$& & $-$ &$-$ &$-$ & $-$ & & $0.04$& $0.93$ & $85.30$& $121.81$ \\[0.1cm]
        
     \rowcolor{burlywood}   \aprone (GraphRNN) & $0.04$  & $92.49$ & $41.42$ &$25.5$& & $0.12$ & $1268.49$ & $211.58$ &$50.41$ & & $0.09$ & $\textbf{6.72}$ & $46.86$ &$5.08$& & $\textbf{0.002}$ & $0.12$ & $41.10$ &$8.86$& & $ \textbf{0.007}$ & $1.05$ & $76.50$ & $2.99$& & $-$ &$-$ & $-$&$-$& & $0.07$  & $1.74$ & $67.10$& $4.7$ \\[0.1cm]
      \rowcolor{burlywood}   \aprtwo (GraphRNN) & $0.06$  & $116.28$ & $190.87$ &$25.5$& &$0.13$  & $796.20$ & $1919.36$ &$50.41$ & & $0.12$  & $8.04$ & $140.63$ &$5.08$& & $\textbf{0.002}$  & $0.08$ &  $204.85$ &$8.86$&  & $0.009$ & $0.97$ & $1266.28$ &$2.99$&  &$-$&$-$ &$-$ &$-$ & & $0.05$ & $1.26$ & $1251.21$& $4.7$  \\[0.1cm]
      \rowcolor{burlywood}   \aprthree (GraphRNN) & $0.05$  & $40.41$ & $152.45$ & $6.33$& & $0.13$ & $764.67$ & $1816.18$ &$77.90$&  & $0.19$ & $12.55$  & $105.59$ & $3.18$& & $0.004$ & $\textbf{0.05}$ & $133.41$&$5.69$& & $\textbf{0.007}$ & $0.85$  & $1250.64$ &$125.4$&  & $-$& $-$&$-$ &$-$ & & $0.04$ & $1.55$ & $1277.53$& $121.81$  \\[0.1cm]
      \rowcolor{vanilla}  \aprone (GRAN) & $\textbf{0.012}$  & $24.75$ & $73.57$& $25.5$&  & $0.12$ & $27974.71$ & $473.26$ & $50.41$& & $\textbf{0.08}$  & $12.81$ & $53.0$ & $5.08$& & $0.005$ & $0.20$  & $47.74$  & $8.86$& & $0.04$ & $6.60$ & $18.75$  & $2.99$& & $\textbf{0.006}$ & $\textbf{6.86}$&$45.56$ &$70.36$& & $0.13$ & $3.68$  & $19.92$& $4.7$  \\[0.1cm]
      \rowcolor{vanilla}  \aprtwo (GRAN) & $0.14$ & $1171.44$  & $175.35$ & $25.5$& & $0.19$  & $15481.15$  & $962.48$ & $50.41$ & & $0.13$  & $45.77$ & $191.80$ &$5.08$& & $0.008$ & $6.90$ &$223.56$ &$8.86$& & $0.008$ & $\textbf{0.56}$ & $846.33$ & $2.99$& & $-$ & $-$ &$-$ & $-$&  & $0.07$ & $2.88$& $843.62$& $4.7$  \\[0.1cm]
      \rowcolor{vanilla}  \aprthree (GRAN) & $0.13$  & $928.87$  & $166.14$ & $6.33$& & $\textbf{0.04}$ & $\textbf{40.6}$ & $1039.26$ & $77.90$& & $0.20$  & $46.03$ & $198.68$ & $3.18$& & $0.03$ & $24.55$ &$206.43$ & $5.69$& & $0.11$  & $9.32$ & $785.09$& $125.4$& & $-$ & $-$ & $-$& $-$ & & $0.13$ & $178.12$ & $835.23$& $121.81$ \\[0.1cm]        
 \bottomrule
   \end{tabular}
    }

    \caption{Comparison of two performance scores for synthetic and real graphs graphs. $|V|_{\mathrm{ave}}$ is the average number nodes and $|E|_{\mathrm{ave}}$ is the average number of edges in the graph set. $\mathrm{t_{tr}}$ indicates the total time to train generative models and $\mathrm{t_{a}}$ is the total time to compute graph alignment using either G-align distance or Fermat distance. ($-$) indicates an out of memory failure. Overall, applying our frameworks to base RNN or base GRAN leads to better performance scores compared to baselines.
    }
    \label{table:4}
\end{table*}

\begin{figure}[!t]
    \centering
    \begin{subfigure}[t]{0.49\linewidth}
        \centering
        \includegraphics[width=.99\linewidth]{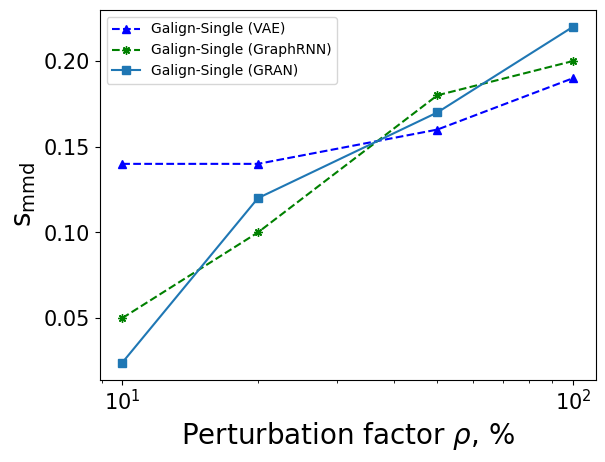}
        \caption{}
        \label{fig:s1-main}
    \end{subfigure}
    \hfil
    \begin{subfigure}[t]{0.49\linewidth}
        \centering
        \includegraphics[width=.99\linewidth]{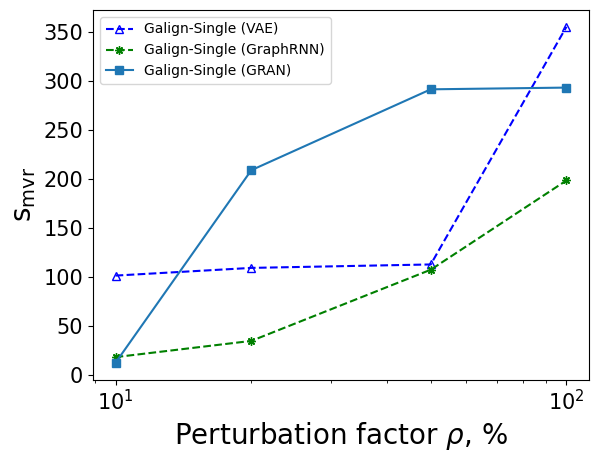}
        \caption{.}
        \label{fig:s2-main}
    \end{subfigure}
    \caption{Sensitivity of $3$ base generative models combined with \aprone \ to noise for community graphs with $|V|= 45$. x-axis: the percentage of edges perturbed, y-axis : $\mathrm{s}_{\mathrm{mmd}}$ (left), $\mathrm{s}_{\mathrm{mvr}}$ (right). \aprone \ (GraphRNN) shows better overall performance, however, in all models there is a direct relation between the perturbation factor and the performance drop.}
    \label{fig:s_noise}
\end{figure}

\noindent\textbf{Impact of Graph Perturbation.}
 We investigate the impact of graph perturbation on the performance of our models by perturbing edges in the $3$-community graphs dataset with $|V|= 45$. The perturbation factor $\rho$ is defined as the percentage of edges that we randomly remove and re-add u.a.r. The $\rho$ values are set to $[10, 20, 50, 100]$ in our experiments. We note that with $\rho <20\%$, graphs still have community structures. In the extreme however, with a $\rho = 100\%$, graphs are effectively Erd\"{o}s-R\'{e}nyi and, thus, their statistics differ significantly from those of the test set.  We compute $\mathrm{s}_{\mathrm{mmd}}$ and $\mathrm{s}_{\mathrm{mvr}}$ for graphs generated with these perturbation factors. 
Fig.~\ref{fig:s_noise} illustrates the performance of the \aprone \ model using GraphRNN, GRAN and VAE as base generative models. \aprone \ (GraphRNN) and \aprone \ (GRAN) models have relatively good $\mathrm{s}_{\mathrm{mmd}}$ compared to \aprone \ (VAE) when $\rho < 50 \%$. At $\rho = 10\%$, \aprone \ (GRAN) has the best performance which is exactly inline with the results we have in Table~\ref{table:4}. As the noise increases, \aprone \ (GraphRNN) shows more robustness to noise compared to the other two models. As expected, all  models are adversely affected when $\rho > 50\%$. 

\section{Conclusion}
We present a group of models that learn distributions of  graphs. Our method is generic with respect to the generative model employed, performs better than the competitors, and enhances permutation invariant and robustness to noise. 

\section*{Acknowledgments}

The authors gratefully acknowledge support by the National Science Foundation (grants  IIS-1741197, CCF-1750539) and Google via GCP
credit support.

\bibliographystyle{IEEEtran}
\bibliography{reference}


\fullversion{}{
\appendix
\section{Alternating Minimization.} \label{sec: am}
At each iteration $t \in \mathbb{N}$, we update $A_0$ and  $\{P_{i}\}_{i \in [n]}$ as follows:

\subsection{Updating \texorpdfstring{${A_0}$}{A0}.} Given that $\{P_{i}\}_{i \in [n]}$ is fixed and $D=0$, minimizing Eq.~\eqref{eq:fermat} w.r.t ${A_0}$ leads to the following problem:
\begin{align}
\begin{split}
 & \underset{A_0\in \mathbb{R}^{m \times m}}{\min } \ 
   \sum_{i=1}^n \|A_iP_{i}^{(t-1)}-P_{i}^{(t-1)}A_0^{(t)} \|
  \label{eq:L4-stp1}
    \end{split}
\end{align}
This problem is convex and at step $t \in \mathbb{N}$ can be solved via convex optimization. Once we solve this optimization problem, we set a threshold to binarize the elements of ${A_0}$. 

\subsection{Updating \texorpdfstring{$\{P_{i}\}_{i \in [n]}$}{Pi}.} Given that $A0$ is fixed and $D= 0$, let $L_P(\{P_{i}\}_{i \in [n]}^{(t)})$ be the loss function at step $t \in \mathbb{N}$. 
\begin{equation}
{\textstyle L_P(\{P_{i}\}_{i \in [n]}^{(t)}))= \sum_{i=1}^n \|A_iP_{i}^{(t)}-P_{i}^{(t)}A_0^{(t)} \| }
 \label{eq:L4-stp2}
\end{equation}
Minimizing Eq.~\eqref{eq:fermat} w.r.t $\{P_{i}\}_{i \in [n]}$ leads to the following problem.

\begin{align}
\begin{split}
 & \underset{ P_{i} \in \mathcal{W}^m}{\min} \ 
  L_P(\{P_{i}\}_{i \in [n]}^{(t)}) 
  \label{eq:L4-stp3}
    \end{split}
\end{align}
This step is convex. It can be solved via optimization toolboxes such as CVXPY~\cite{diamond2016cvxpy}  or efficient algorithms such as Frank-Wolfe algorithm~\cite{frank1956algorithm}. Frank-Wolfe algorithm is explained in details in the Section~\ref{app:fw}.

\section{Frank Wolfe.}
\label{app:fw}
The objective function in Eq.~\eqref{eq:L4-stp2} can be solved via Frank-Wolfe algorithm. Frank-Wolfe is an iterative algorithm that solves the problem through a sequence of linear programs (LPs). This algorithm starts from a feasible ${P}^0 \in \mathcal{W}^m$, e.g. , the identity  matrix ${I}$ and in each iteration $t \in \mathbb{N}$ proceeds as follows:

\begin{subequations}
\begin{align}
 &S^{(t)}=  \underset{ S_{ij} \in \mathcal{W}^m, S_{ii}=I, S \succeq 0}{\mathrm{arg} \ \min} \ \trace(S^T, {\nabla}_{{{P}}}L_P({{P}}^{(t)}))  
  \label{eq:L62}\\
 &  {{P}}^{(t+1)} = (1-{\gamma}_t) {P}^{(t)}+{\gamma}_t S^{(t)},
  \label{eq:L63}
\end{align}
\end{subequations}
where ${\gamma}_t$ is the step size and can be set to e.g. $\frac{2}{t+2}$ or determined by line search~\cite{boyd2004convex} as follows: 

    \begin{equation}
       {\gamma}_t= {\mathrm{arg} \ \min_{{\gamma}_t \in [0, 1]}} L_P(((1-{\gamma}_t){P}^{(t)}+{\gamma}_t S^{(t)})
       \label{eq:gam2}
    \end{equation}


\section{Table of metrics.}  \label{app:met}  
In the Table~\ref{table:2} we provide the lists of metrics we measured in the experiments and their description. 
\begin{table}[ht]

    \centering
    \begin{scriptsize}
    \begin{tabular}{c l}
        \toprule
        Notation & Description \\
        \midrule
        \degdist & Graphs degree distribution\\

        \clcoef & distribution of clustering coefficient of nodes \\ 
         & for each graph in the graph set\\ 
        \aso:& assortativity, Pearson correlation coefficient of \\
        & degree between pairs of linked nodes\\
       \tri: & number of triangles for each graph in the graph set\\
   
       \wc: & wedge count,  number of wedges for each graph\\
        &  in the graph set\\

        \clc: & claw count, number of claws for each graph \\
        & in the graph set\\
        \bottomrule
        \end{tabular}
    \end{scriptsize}

    \caption{Summary of metrics.}
    \label{table:2}

\end{table}

 \section{Performance scores.}  \label{perf}
In order to calculate ${\mathrm{MMD}}^2$ , let a function $f$ belong to a unit ball in a reproducing kernel Hilbert space (RKHS) $\mathcal{H}$, $f \in \mathcal{H}$, and $k$ be the kernel. The ${\mathrm{MMD}}^2$ between two sets of samples $\{x_i\}_{i=1}^N {\sim}^{\mathrm{iid}} p$ and  $\{y_i\}_{i=1}^N {\sim}^{\mathrm{iid}} q$ from distributions $p$ and $q$ is computed as follows:

\begin{fleqn}
\begin{equation}
\begin{aligned}[b]
    {\mathrm{MMD}}^2 &= \\
     & \frac{1}{N(N-1)} {\sum}_{i=1}^N {\sum}_{j \neq i}^N (k(x_i, x_j)+ k(y_i, y_j)) \\
    &  -\frac{1}{N^2}  {\sum}_{i=1}^N {\sum}_{j=1}^N (k(x_i, y_j)+k(x_j, y_i))
\end{aligned}
\end{equation}
\end{fleqn}

The performance of ${\mathrm{MMD}}^2$ depends on choice of the kernel. Here we use Gaussian-Wasserstein RBF kernel $k(x, y)= e^{-\frac{W(p, q)^2}{2{\sigma}^2}}$ , where $W(p, q) $ is the first Wasserstein distance. The $k(x, y)$ function is bounded, $k(x, y) \in [0, 1] $ and therefore ${\mathrm{MMD}}^2 \in [0, 2]$.

\noindent\textbf{$s_{\mathrm{mmd}}$ score.} Combining ${\mathrm{MMD}}^2$ of all metrics we measured, we present $s_{\mathrm{mmd}}$ score to assess the overall quality of generated graphs.

\begin{fleqn}
\begin{equation}
\begin{aligned}[b]
\mathrm{s}_{\mathrm{mmd}} & = \\ &\frac{1}{12}({\mathrm{MMD}}^2(\mathrm{\degdist})+  {\mathrm{MMD}}^2(\mathrm{\clcoef}) + {\mathrm{MMD}}^2(\mathrm{\aso}) \\
&  + {\mathrm{MMD}}^2(\mathrm{\tri})+ {\mathrm{MMD}}^2(\mathrm{\wc})+ {\mathrm{MMD}}^2(\mathrm{\clc}))
\label{eq:s1}
\end{aligned}
\end{equation}
\end{fleqn}

Note that $\mathrm{s}_{\mathrm{mmd}} \in [0, 1]$. The smaller this score, the smaller the distance between the generated graphs and test set.

\noindent\textbf{$s_{\mathrm{mvr}}$ score.} Our second performance score, $\mathrm{s}_{\mathrm{mvr}}$ is formulated as follows:
\begin{sloppypar}
\begin{fleqn}
\begin{equation}
\begin{aligned}[b]
\mathrm{s}_{\mathrm{mvr}} &= \frac{1}{6}(\\  &\frac{(\mu_{\mathrm{r}}(\degdist)-\mu_{\mathrm{g}}(\degdist))^2}{{\sigma}^2_{\mathrm{r}}(\degdist)}
\\
 &  +\frac{(\mu_{\mathrm{r}}(\aso)-\mu_{\mathrm{g}}(\aso))^2}{{\sigma}^2_{\mathrm{r}}(\aso)}\\ 
& + \frac{(\mu_{\mathrm{r}}(\clcoef)-\mu_{\mathrm{g}}(\clcoef))^2}{{\sigma}^2_{\mathrm{r}}(\clcoef)} \\
 &    + \frac{(\mu_{\mathrm{r}}(\tri)-\mu_{\mathrm{g}}(\tri))^2}{{\sigma}^2_{\mathrm{r}}(\tri)}\\
 &      + \frac{(\mu_{\mathrm{r}}(\wc)-\mu_{\mathrm{g}}(\wc))^2}{{\sigma}^2_{\mathrm{t}}(\wc)}
     + \\
 &   \frac{(\mu_{\mathrm{r}}(\clc)-\mu_{\mathrm{g}}(\clc))^2}{{\sigma}^2_{\mathrm{r}}(\clc)}),
\label{eq:s2}
\end{aligned}
\end{equation}
\end{fleqn}
\end{sloppypar}

\fussy where $\mu_{\mathrm{r}}$, $\mu_{\mathrm{g}}$ and ${\sigma}^2_{\mathrm{t}}$ represent mean value for the reference set, mean value for the generated set and variance of the reference set, respectively.

\section{Accelerating Multi-distances} \label{app:acmd}
\fussy In Alg.~\ref{alg:cap1} we explain how to compute the final center graph by G-Parallel. We describe C-Serial algorithm is in details in Alg.~\ref{alg:cap2} and the detail of CG-Parallel  are in Alg.~\ref{alg:cap3}. Table~\ref{tab:tab_data} shows summary of datasets, acceleration methods and solvers used to compute graph alignment. 
 \begin{algorithm}[!t]
 
\scalebox{0.7}{
\begin{minipage}{\linewidth}
\caption{G-Parallel: Grouping and Parallelizing Graphs}\label{alg:cap1}
\KwInput{$\bm{\mathcal{G}}=\{\mathcal{G}_1,\mathcal{G}_2,...,\mathcal{G}_n\}$, $K:$ number of graphs in each group.} 
\KwOutput{$\mathcal{G}_{0_{\mathrm{out}}}:$ the center graph.}
\For {$k= \{ 0, 1, 2, \cdots, [\frac{n}{K}]\}$}{$\bm{\tilde{\mathcal{G}_k}}= \{\mathcal{G}_{1+k\times K}, \mathcal{G}_{2+k\times K}, \cdots, \mathcal{G}_{K+k\times K} \}$

$\bm{\tilde{\mathcal{G}_k}}=\mathrm{align}(\bm{\tilde{\mathcal{G}_k}})$

${\tilde{\mathcal{G}_0}}^k= \mathrm{center}(\bm{\tilde{\mathcal{G}_k}})$}
$\bm{\tilde{\mathcal{G}}}= \{{\tilde{\mathcal{G}_0}}^k, \  \mathrm{for} \ k= \{ 0, 1, 2, \cdots, [\frac{N}{K}]\} \} $

\If {$[\frac{n}{K}] > K$}
{
\While{$[\frac{n}{K}] > K$}
{$n= [\frac{n}{K}]$

$\bm{\mathcal{G}}= \bm{\tilde{\mathcal{G}}}$

\For {$k= \{ 0, 1, 2, \cdots, [\frac{n}{K}]\}$}{$\bm{\tilde{\mathcal{G}_k}}= \{\mathcal{G}_{1+k\times K}, \mathcal{G}_{2+k\times K}, \cdots, \mathcal{G}_{K+k\times K} \}$

$\bm{\tilde{\mathcal{G}_k}}=\mathrm{align}(\bm{\tilde{\mathcal{G}_k}})$

${\tilde{\mathcal{G}_0}}^k=\mathrm{center}(\bm{\tilde{\mathcal{G}_k}})$}
$\bm{\tilde{\mathcal{G}}}= \{{\tilde{\mathcal{G}_0}}^k \mathrm{for} \ k= \{ 0, 1, 2, \cdots, [\frac{n}{K}]\} \} $

}
}

 $\bm{\mathcal{G}}_{\mathrm{out}}= \bm{\tilde{\mathcal{G}}}$
 
$\mathcal{G}_{0_{\mathrm{out}}}= \mathrm{center}(\bm{\mathcal{G}}_{\mathrm{out}})$
 \end{minipage}%
    }
\end{algorithm}

 \begin{algorithm}[!t]
\scalebox{0.7}{
\begin{minipage}{\linewidth}
\caption{C-Serial: Coarsening Graphs}\label{alg:cap2}
\begin{small}
\KwInput{$\bm{\mathcal{G}}=\{\mathcal{G}_1,\mathcal{G}_2,...,\mathcal{G}_n\}$, $c:$ number of clusters in graphs.} 
  \KwOutput{$\mathcal{G}_{0_{\mathrm{out}}}:$ the center graph.}
  
\For {$l= \{ 0, 1, 2, \cdots, n\}$}
{$\hat{G}_{l}= \mathrm{coarsen}(G_{l})$}
$\bm{\hat{\mathcal{G}}}= \mathrm{align}(\{\hat{\mathcal{G}}_{1}, \hat{\mathcal{G}}_{2}, \cdots, \hat{\mathcal{G}}_{n} \})$\\
\For {$l= \{ 0, 1, 2, \cdots, n\}$}
{
align clusters given their alignment in the coarsened graphs.
}

compute center of clusters.

compute center of edges connecting clusters.\\

${\tilde{\mathcal{G}_0}}^k$: build given the center of clusters and center of edges connecting clusters.

\end{small}
 \end{minipage}%
    }
\end{algorithm}

 \begin{algorithm}[!t]
\scalebox{0.7}{
\begin{minipage}{\linewidth}
\caption{CG-Parallel: Coarsening, Grouping and Parallelizing Graphs}\label{alg:cap3}
\begin{small}
\KwInput{$\bm{\mathcal{G}}=\{\mathcal{G}_1,\mathcal{G}_2,...,\mathcal{G}_n\}$, $K:$ number of graphs in each group, $c:$ number of clusters in graphs.} 
  \KwOutput{$\mathcal{G}_{0_{\mathrm{out}}}:$ the center graph.}
  
\If {$[\frac{n}{K}] > K$}
{\While{$[\frac{n}{K}] > K$}
{
\For {$k= \{ 0, 1, 2, \cdots, [\frac{n}{K}]\}$}{$\bm{\tilde{\mathcal{G}_k}}= \{\mathcal{G}_{1+k\times K}, \mathcal{G}_{2+k\times K}, \cdots, \mathcal{G}_{K+k\times K} \}$\\
\For {$l= \{ 0, 1, 2, \cdots, K\}$}
{$\hat{G}_{l+k \times K}= \mathrm{coarsen}(G_{l+k \times K})$}
$\bm{\hat{\mathcal{G}_k}}= \mathrm{align}(\{\hat{\mathcal{G}}_{1+k\times K}, \hat{\mathcal{G}}_{2+k\times K}, \cdots, \hat{\mathcal{G}}_{K+k\times K} \})$\\
\For {$l= \{ 0, 1, 2, \cdots, K\}$}
 {
align clusters given their alignment in the coarsened graphs.
}

compute center of clusters.

compute center of edges connecting clusters.\\

${\tilde{\mathcal{G}_0}}^k$: build given the center of clusters and center of edges connecting clusters. }
$\bm{\tilde{\mathcal{G}}}= \{{\tilde{\mathcal{G}_0}}^k, \  \mathrm{for} \ k= \{ 0, 1, 2, \cdots, [\frac{N}{K}]\} \} $
}




}

 $\bm{\mathcal{G}}_{\mathrm{out}}= \bm{\tilde{\mathcal{G}}}$
 
$\mathcal{G}_{0_{\mathrm{out}}}= \mathrm{center}(\bm{\mathcal{G}}_{\mathrm{out}})$
\end{small}
 \end{minipage}%
    }
\end{algorithm}

\begin{table}[!t]
    \resizebox{\columnwidth}{!}{
    \begin{scriptsize}
    \begin{tabular}{*{7}{c}}
        \toprule
        & $|V|_{\mathrm{ave}}$ & $|E|_{\mathrm{ave}}$ & $n$ & Alignment alg. & Solver (Fermat) & Solver (G-align) \\
        \midrule
Community (small)& $45$ & $98$ & $100$ & G-Parallel & CVXPY + AM& CVXPY  \\ Community (large)&  $150$&  $2727$& $100$ &  CG-Parallel & CVXPY + AM& CVXPY\\
Grid& $36$& $265$& $100$ & G-Parallel & CVXPY + AM & CVXPY  \\
Ego-Citeseer & $35$& $65$ & $100$  & G-Parallel& CVXPY + AM & CVXPY\\ 
Ego-B-A (small) & $118$& $298$& $100$  & CG-Parallel & CVXPY + AM & CVXPY\\ 
Ego-B-A (large)& $1028$& $1471$ & $68$& CG-Parallel & CVXPY + AM & CVXPY\\
\bottomrule
\end{tabular}
\end{scriptsize}
}
\caption{Dataset summary including average number of nodes and edges and number of graphs in the graph set, along with the algorithm and solvers used to compute graph alignment in Fermat and G-align distance. For smaller graphs (with $|V|_{\mathrm{ave}} < 50$ ) we use the G-Parallel method. For larger graphs, to further accelerate computing the graph alignment we use CG-Parallel. By using these acceleration techniques, all alignment problems can be solved by CVXPY.}
\label{tab:tab_data}
\end{table}

\fussy \section{Implementation Details}
\label{app:impd}
We compared the performance our models against five different deep baseline described below.

\noindent\textbf{GraphRNN.}
 You et al.~\cite{you2018graphrnn} proposes a framework based on graph neural networks. This model uses a graph-level RNN to add a new node to a node sequence each time step and an edge-level RNN to model the generation process of nodes and edges. The reference code for this model is provided by the authors and we followed their recommendation for setting the hyperparameters.

 \noindent\textbf{GRAN.}  Liao et al.~\cite{liao2019efficient} proposes a graph recurrent attention framework. This model uses an attention-based GNN and generates a block of graphs that consists of multiple rows of graph adjacency matrices conditioned on the previously generated blocks of the graph and uses a group canonical node ordering, e.g., DFS and BFS to address node ordering problem.
 
\noindent \textbf{VAE.} Kipf $\&$  Welling~\cite{kipf2016variational} propose a variational autoencoder that is characterized by a probabilistic inference model that maps observed data to a latent representation, a prior distribution over the latent variables and a probabilistic generative model. We randomly pick a graph with a random node ordering from graph set $\bm{\mathcal{G}}$ and train a VAE to generate graphs.

\noindent\textbf{GraphVAE.}
Simonovsky $\&$ Komodakis~\cite{simonovsky2018graphvae} propose a variational autoencoder that outputs a probabilistic
fully-connected graph and uses a graph matching
algorithm to align graph to the ground truth. GraphVAE outputs a graph with adjacency matrix, node attributes and edge attributes. We adapt it to our problem by using one-hot representations of the features. The encoder is a graph convolutional network and the decoder is a multi-layer perception. We used code for this model from \cite{you2018graphrnn} and set the hyperparameters
based on recommendations made in \cite{simonovsky2018graphvae}.

\noindent\textbf{DeepGMG.}
Li et al.~\cite{li2018learning} introduce a generative model for graphs that generates graphs in a sequential manner. It generates one node at a time and
connects each node to the partial graph already generated by creating edges one by one. We used  the implementation in \cite{you2018graphrnn}  and the hyperparameters were set based on the recommendations made in \cite{li2018learning}.

We take $80\%$ of graphs for training and the rest for the test sets. During testing, GraphRNN model and GRAN model generate graphs directly. However, the output of the VAE decoder is an adjacency matrix with elements in the range of $[0, 1]$. We binarize the adjacency matrix by applying a threshold, ${\tau}$. We find ${\tau}$ by comparing two sets of graphs, the ones generated from the VAE and $20\%$ of the graphs in the training set, and computing two scores, which we denote $s_{\mathrm{mmd}}$ and $s_{\mathrm{mvr}}$ (see Sec.~\ref{metrics}) to measure the distance between these two sets for a range of values of $\tau$. We chose the value of $\tau$ that returns the smallest $s_{\mathrm{mmd}}$ as our optimal threshold in testing.
In all our \apr models, we pre-compute the graph alignment for all datasets and use the aligned graphs for training the generative models. We use GraphRNN~\cite{you2018graphrnn}, GRAN~\cite{liao2019efficient} and  VAE~\cite{kipf2016variational} as our base generative models. We followed the instructions given in~\cite{you2018graphrnn} and ~\cite{liao2019efficient} to set the hyperparameters in GraphRNN~\cite{you2018graphrnn} and GRAN~\cite{liao2019efficient} 
and for the VAE~\cite{kipf2016variational} we used the hyperparameters given in~\cite{kipf2016variational} and set the learning rate to $0.001$. (We note that VAE\cite{kipf2016semi} here refers to the model proposed by Kipf $\&$ Welling and is different from GraphVAE\cite{simonovsky2018graphvae}  by Simonovsky $\&$ Komodakis. In all models, the hidden dimensions $\{Z_i\}_{i \in [n]}$ of small graphs are set to $16$. For medium graphs ($|V| \in [100- 500]$ ) the hidden dimensions are $64$ and the hidden dimensions of the large graphs ($|V|> 500$) are set to $256$. In all experiments, The node features are one-hot indicator vectors.
The \apr \ architectures are implemented in Python3 using Tensorflow and Pytorch. We implemented the solution of the constrained optimization problems in Section ~\ref{sec:method} via CVXPY. We implemented all solvers in Python3. For clustering graphs we use Scikit-learn~\cite{scikit-learn}. All experiments are carried out on a Tesla V100 GPU with
$32$ GB memory and $5120$ cores. G-Parallel and CG-Parallel methods parallelizes the computations using python multiprocessing package. For both of these parallel graph alignment algorithms we use a single machine with $40$ CPUs.

}

\end{document}